\documentclass[
  twocolumn,
  prl,
  showpacs,
  amsmath,
  amssymb,
  superscriptaddress,
  floatfix
]{revtex4-1}

\usepackage{bm}
\usepackage{graphicx}
\usepackage{multirow}
\usepackage[dvips]{color}

\newcommand{\be}{\begin{equation}}
\newcommand{\ee}{\end{equation}}
\newcommand{\bea}{\begin{eqnarray}}
\newcommand{\eea}{\end{eqnarray}}
\newcommand{\ba}{\begin{array}}
\newcommand{\ea}{\end{array}}

\newcommand{\pc}[1] {p_{\rm c}^{#1}}

\begin{document}

\title{Recursive percolation}

\author{Youjin Deng}
\email{yjdeng@ustc.edu.cn}
\affiliation{Hefei National Laboratory for Physical Sciences at Microscale
and Department of Modern Physics, \\
University of Science and Technology of China, Hefei, Anhui 230026, China}

\author{Jesper Lykke Jacobsen}
\email{jesper.jacobsen@ens.fr}
\affiliation{Laboratoire de Physique Th\'eorique, \'Ecole Normale Sup\'erieure, 24 rue Lhomond,
  75231 Paris, France}
\affiliation{Universit\'e Pierre et Marie Curie, 4 place Jussieu,
  75252 Paris, France}

\author{Xuan-Wen Liu}
\affiliation{Hefei National Laboratory for Physical Sciences at Microscale
and Department of Modern Physics, \\
University of Science and Technology of China, Hefei, Anhui 230026, China}

\begin{abstract}
We introduce a simple lattice model in which percolation is constructed on top of critical percolation clusters, 
and show that it can be repeated recursively any number $n$ of generations.
In two dimensions, we determine the percolation thresholds up to $n=5$. 
The corresponding critical clusters become more and more compact as $n$ increases,
and define universal scaling functions of the standard two-dimensional form 
and critical exponents that are distinct for any $n$.
This family of exponents differs from any previously known universality class, 
and cannot be accommodated by existing analytical methods. 
We confirm that recursive percolation is well defined also in three dimensions. 
\end{abstract}


\maketitle

The use of percolation theory pervades many parts of science, ranging from material science to geology,
epidemiology and sociology \cite{StaufferAharony,Grimmett}.
At the percolation thres\-hold it leads to random, scale invariant
geometries that have become paradigmatic in theoretical physics and probability theory. Lattice models
for percolation \cite{Hammersley} have propelled powerful theoretical constructions, leading to a host of
exact results, particularly in two dimensions \cite{CFT,loopreview,Schramm,KagerCardy}.

A typical model is bond percolation, in which each link of the lattice is taken to be open with probability $p$.
An important assumption in this model is that the medium is independent of its preceding history.
However, in nume\-rous situations this hypothesis is not fulfilled. 
Examples include the percolation of a liquid in a porous medium like granular rocks~\cite{Herrmann04}, or
epidemic spread~\cite{Tome2010,Sneppen2010},
where a renewed percolation (resp.\ spread) event may depend on the history of sedimentation
(resp.\ immunization).
In both cases, the first percolation process imposes a particular type of quenched disorder 
on the following process.
The purpose of this Letter is to formulate a simple model
of such {\em recursive percolation} and study its properties numerically.

Given a configuration of percolation clusters at criticality, $p^0 = \pc{0}$, with 
superscript $n=0$ for the original percolation, we define a new ($n=1$) percolation 
process on top of them such that occupied bonds are placed with probability $p^1$ on 
all the pairs of neighboring sites in the same cluster. 
One might then expect that any finite probability $p^1 < 1$ would destroy 
the critical singularity and lead to a subcritical phase where it becomes exponentially difficult to form a large cluster. 
Contrary to this expectation, we show that there exists a non-trivial critical threshold, $1> \pc{1} > \pc{0}$, 
separating a subcritical and a critical phase.
This means in particular that the construction can be repeated recursively: on top of
the new critical clusters, one may again study a percolation process and search for its
threshold. The same scenario takes place, so that the construction may be repeated any number of times. 
Surprisingly, the $n$th generation of percolation clusters
thus generated enjoys, at their threshold $p = \pc{n}$, distinct critical exponents for any $n$. 
The $d=2$ simulations show that the exponents are universal, i.e., independent of lattice 
and percolation process (bond/site).  Moreover, they tend to finite limits
when $n \to \infty$---in the case of a ``worn out'' medium. 

This family of recursive critical exponents for $n \ge 1$
does not appear in any previously known universality class \cite{CFT}.
Neither can the $d=2$ exponents be accounted for by existing analytical constructions,
including the Coulomb gas (CG) approach to conformal field theory (CFT) \cite{CFT,loopreview},
and the more recent Schramm-Loewner evolution (SLE) \cite{Schramm,KagerCardy}. 
These field-theoretical methods have provided a plenitude of information about critical behavior,
predicting exact values \cite{NRS80,SD87} of critical exponents for most two-dimensional lattice models.
We find in particular that recursive percolation for $n \geq 1$ 
violates the domain Markov property in the context of SLE theory.

\paragraph{Percolation threshold.}
We study recursive percolation on periodic $L \times L$ square
lattices. The starting point is standard bond percolation~\cite{StaufferAharony,Grimmett}, 
with the known thres\-hold $\pc{0} = \frac12$. From a given set
of percolation clusters ${\cal C}_0$, henceforth called {\em standard clusters} for clarity,
we define a set of {\em dense clusters} $\overline{\cal C}_0$ by filling in all bonds
between neighboring sites in the same cluster.
Here and elsewhere quantities with (resp.\ without) an overline refer to the dense 
(resp.\ standard) case. 

Suppose that the thres\-holds $\pc{1},\ldots,\pc{n-1}$ are already known. A configuration of
clusters ${\cal C}_n$ at generation $n \ge 1$, with a given occupation probability $p^n$, is then
defined as follows:
For each $i=1,\ldots,n$ in turn, produce ${\cal C}_i$ by performing bond percolation on
$\overline{\cal C}_{i-1}$ with probability $p^i = \pc{i}$ if $i < n$, and $p^i = p^n$ if $i = n$.

\begin{figure}
\centering
\vspace{0.5cm}
\hspace*{-1.0cm}
\includegraphics[scale=0.75]{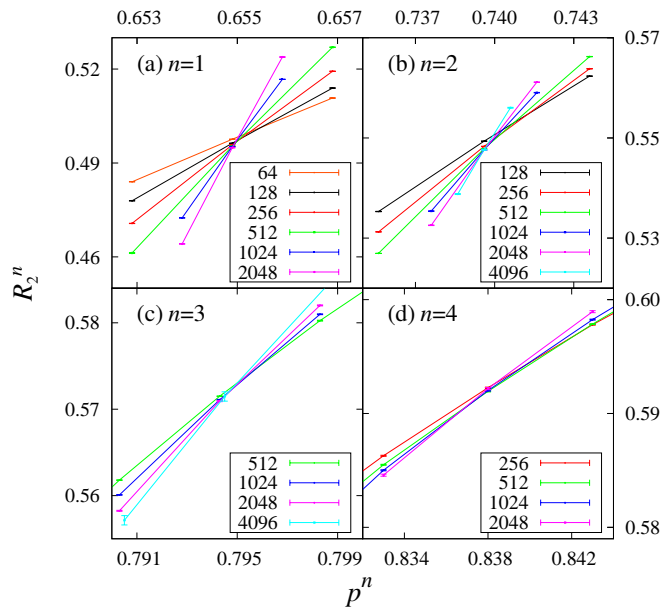}
\vspace{0.5cm}
\caption{(color online) Wrapping probability $R_2^n$ versus $p^n$ for $n=1,2,3,4$.
         The superscript $n$ is for the $n$th generation.
         Percolation threshold is located by the approximately common crossing 
         for different sizes $L$, as denoted by different colors.}
\label{fig1_R2}
\end{figure}

We have performed extensive simulations for $L = 2^\ell$, with $\ell=4,5,\ldots, 12$.
The existence of a non-trivial threshold $\pc{n}$ is revealed by the crossing properties of the
probability $R_2^n$ that one cluster in ${\cal C}_n$ wraps both periodic lattice directions
(see Fig.~\ref{fig1_R2}).
The finite-size scaling clearly shows that $\pc{n}$ acts as an unstable fixed point for
$n$th generation clusters, sustaining flows to the trivial fixed points $p^n \approx 1$
and $p^n = 0$ respectively; see Supplemental Material (SM) for more details.

\begin{table}[htpb]
\centering
\caption{Threshold $\pc{n}$, exponent $y_t^n$, and wrapping probability $R_2^n$ for $n$th generation percolation.}
\begin{tabular}{cllllll}
\hline
$n$        & 1           & 2          & 3          & 4          & 5 \\
\hline
$y_t^{n}$  & 0.433(1)    & 0.273(4)   & 0.182(4)   & 0.116(10)  & 0.09(2)  \\
$R_2^n$    & 0.495(1)    & 0.547(1)   & 0.571(2)   & 0.586(3)   & 0.595(4) \\
$\pc{n}$   & 0.654902(10) & 0.73954(4) & 0.7945(1)  & 0.8342(8)  & 0.861(4) \\
\hline
\end{tabular}
\label{tab:1}
\end{table}

From the scaling of $R_2^n$ near $\pc{n}$, we have determined, for $n \le 5$,
the thresholds $\pc{n}$, the thermal exponent $y_t^n$, 
and the critical value of $R_2^n$ (see Tab.~\ref{tab:1}).
Notice that the values of $\pc{n}$ are close to the
simple fraction $(n+1)/(n+2)$, especially for larger $n$. This suggests
that $\pc{n} \to 1$ for $n \to \infty$, meaning that recursive percolation can be defined for
any number of generations. 

\begin{figure}
\centering
\hspace*{-0.1cm}\includegraphics[scale=0.35]{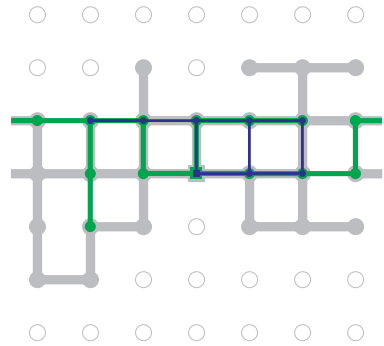}
\caption{(color online) Illustration of recursive single-cluster growing processes.
 The $n=0, 1, 2$ bonds and clusters are marked in gray, green, and black, respectively.
} 
\label{fig2_recper}
\end{figure}

\paragraph{Observables and scaling.}
\begin{table}[htpb]
\centering
\caption{Values of critical exponents. For $n=0$, the backbone is estimated as 
$d_{\rm B}^0 = 1.6431(6)$ from a transfer matrix computation \cite{Jacobsen2002} 
and $d_{\rm B}^0 = 1.643\,36(10)$ by Monte Carlo simulations \cite{Deng2004,Zhou2012}.
Note that $d_{\rm R}^n$ coincides with $y_t^n$ in Tab.~\ref{tab:1}.
The equality $y_t^0 = d_R^0$ holds true in any spatial dimension~\cite{Coniglio82,Vasseur12}. 
}
\begin{tabular}{llllll}
\hline
$n$                         &\;0         &\;1         &\;2          &\;3          &\;4 \\
\hline
$d_{\rm F}^n$               &\;1.8958(1) &\;1.8573(1) &\;1.8424(1)  &\;1.8357(2)  &\;1.8323(2) \\
$d_{\rm B}^n$               &\;1.6433(3) &\;1.7596(1) &\;1.7942(1)  &\;1.8078(2)  &\;1.8148(2) \\
\hline
$d_{\rm H}^n$               &\;1.75      &\;1.6083(1) &\;1.5358(1)  &\;1.4967(1)  &\;1.4723(2) \\
$\overline{d}_{\rm H}^n$    &\;1.3333    &\;1.3739(1) &\;1.3929(1)  &\;1.4026(2)  &\;1.4075(2) \\
\hline
$d_{\rm R}^n$               &\;0.751(1)  &\;0.433(1)  &\;0.272(2)   &\;0.182(2)   &\;0.121(3) \\
$\overline{d}_{\rm R}^n$    &-0.77(3)    &-0.429(1)   &-0.275(2)    &-0.194(6)    &-0.15(1) \\
\hline
\end{tabular}
\label{tab:4}
\end{table}
We also measured the size $C_1^n$ of the largest cluster, 
the number $B_{\rm R}^n$ of pseudo-bridges, the length $H_1^n$ of the largest loop 
surrounding percolation clusters, and the size $C_{\rm b1}^n$ of the largest backbone 
clusters (see SM for detailed definitions).
At criticality, the finite-size scaling of these observables is 
governed by a set of critical exponents, 
 \begin{equation}
 C_1^n \propto L^{d^n_{\rm F}} \; ,   \hspace{1mm}  
 H_1^n \propto L^{d^n_{\rm H}} \; ,      \hspace{1mm} 
 B_{\rm R}^n \propto L^{d^n_{\rm R}} \; , \hspace{1mm} 
 C_{\rm b1}^n \propto L^{d^n_{\rm B}} \; , 
 \label{eq:scaling}
 \end{equation}
where $d_{\rm F}^n$ is the cluster's fractal dimension, $d_{\rm H}^n$ is  
the hull dimension, $d_{\rm R}^n$ is the red-bond exponent, 
and $d_{\rm B}^n$ is the backbone dimension. 
These critical exponents characterize more precisely the critical clusters ${\cal C}_n$. 
Analogous measurements were taken for the dense clusters $\overline{\cal C}_n$, 
and our definitions imply that $d_{\rm F}^n = \overline{d}_{\rm F}^n$.
The ``dense" hulls correspond to the accessible perimeters in Ref.~\cite{GrossmanAharony}.

The scaling behavior in Eq.~(\ref{eq:scaling}) is well confirmed by our numerical data,
and the results are shown in Tab.~\ref{tab:4}~\cite{ErrorBar}. 

\paragraph{Scaling functions.}
\begin{figure}
\centering
\hspace*{-0.1cm}\includegraphics[scale=0.65]{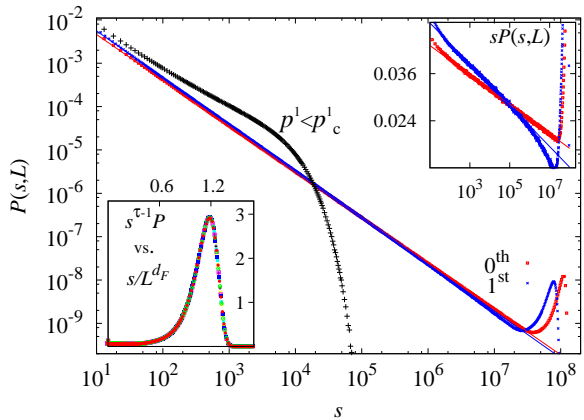}
\caption{ (color online) Probability distribution $P(s,L)$ in the single-cluster growing procedure.
        The size $L=16 \, 384$ in the main plot.
        The red ($n=0$) and blue ($n=1$) data are for $p^0 = \pc{0} = 1/2$ and $p^1 = \pc{1} = 0.654902$,
        respectively. 
        The black curve with $p^1=0.6 <p_c^1$ displays sub-critical behavior.
        The red/blue straight lines have slopes $1-\tau \equiv -d/d_{\rm F}^n$,
        with $d_{\rm F}^0=91/48$ and $d_{\rm F}^1=1.8573$.
        The right-top inset shows the product $s P(s,L)$, 
        and the left-bottom corner displays $s^{\tau-1} P(s,L)$ for $n=1$ versus $s/d_{\rm F}^1$, 
        with $L=256, 512, \ldots, 16\,384$. 
  }
\label{fig3_histo}
\end{figure}
The recursive percolation can be constructed in an alternative way: 
Start from a seed site, grow a percolation cluster, 
construct a $n=1$ cluster right on top of it from the same seed site, 
and repeat the process recursively. 
This is illustrated in Fig.~\ref{fig2_recper}.
We employ this procedure on periodic $L \times L$ square lattices, and
record the probability distribution $P(s,L)$ that the grown cluster is of size $s$. 
Figure~\ref{fig3_histo} shows $P(s,L)$ at criti\-cality versus $s$ in a log-log scale for $L=16\,384$.
The algebraically decaying behavior of $P(s,L)$ is well displayed in a wide range of size $s$.

The standard scaling theory yields
\begin{equation}
P(s,L) \sim s^{1-\tau} f(s/L^{d_{\rm F}}) \; ,  \hspace{5mm} (\tau = 1+d/d_{\rm F}) 
\label{eq:scaling_P}
\end{equation}
where $f$ is a universal function and the hyperscaling relation $\tau = 1+d/d_{\rm F}$ 
involves spatial dimension $d$.
A non-trivial question arises: Does Eq.~(\ref{eq:scaling_P}), 
particularly the hyperscaling relation, hold true for $n \geq 1$, 
for which the underlying geometries are already fractal?
We apply in Fig.~\ref{fig3_histo} the critical exponents $d_{\rm F}^0=91/48$ \cite{NRS80}
and $d_{\rm F}^1=1.8573$. 
The latter is taken from Tab.~\ref{tab:4}, 
obtained from the other construction of recursive percolation.
Surprisingly, the two insets of Fig.~\ref{fig3_histo} strongly support that 
the $n=1$ recursive percolation enjoys the scaling form in Eq.~(\ref{eq:scaling_P}) with original dimensionality $d=2$.

\begin{figure}
\centering
\vspace{0.5cm}
\hspace*{-1.1cm}\includegraphics[scale=0.75]{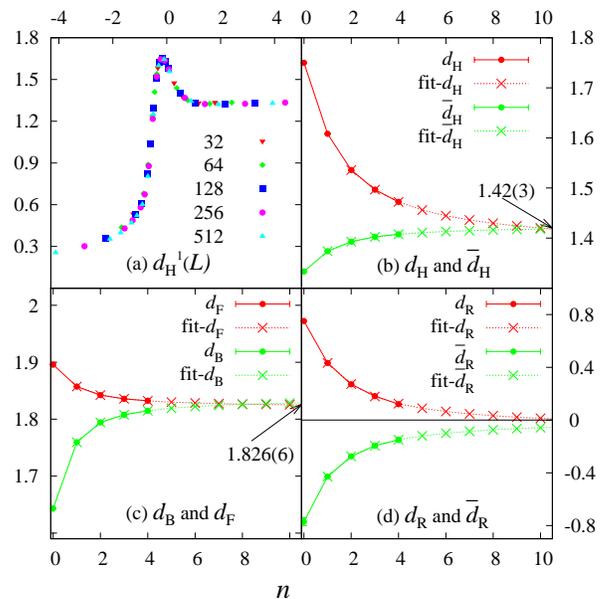}
\vspace{1cm}
\caption{(color online) Fractal dimensions. (a) effective dimensions $d_{\rm H}^1(L)$ versus 
 the variable $u = (p^1-\pc{1}) L^{y_t^1}$; $d_{\rm H}^1(L)$ approaches the exact value $4/3$ for $u >>0$.
  (b)-(d) the critical exponent $d_{\rm X}^n$
 versus generation $n$, with X $=$ H, F, R, respectively.
 The ``$\times$" points are from the fitting results.
 For each of the exponents, the $n$-dependence can be fitted to a common $n \to \infty$ limit as shown.
}
\label{fig4_dX}
\end{figure}

We show in Fig.~\ref{fig4_dX}(a) the effective hull dimension
$d_{\rm H}^1$ against the variable $u = (p^1-\pc{1}) L^{y_t^1}$,
defined as $d_{\rm H}^1 (L) \equiv  \log_2 [H_{1}^1(2L)/H_{1}^1(L)] $.
With the choice $y_t^1 = 0.433(1)$ the data for all sizes $L$ collapse perfectly to reveal
the universal scaling function. For $u=0$ one has the ${\cal C}_n$ universality class,
here with $d_{\rm H}^1 = 1.6083(1)$, while for $u>0$ there is a flow to the
$\overline{\cal C}_{n-1}$ universality class, as expected, with now
$\overline{d}_{\rm H}^0 = 4/3$. The flow for $u<0$ is to the trivial fixed point with $d_{\rm H} = 0$.

\paragraph{Critical exponents.}
In two dimensions, the field-theoretical methods \cite{NRS80,SD87} predict 
the exact results for percolation and $q$-state Fortuin-Kasteleyn (FK) clusters (a correlated percolation model) 
\begin{align}
&d_{\rm F}^0 = 2-(6-g)(g-2)/(8g) = 91/48 \,, \notag\\
&d_{\rm H}^0 =1+2/g  = 7/4 \,, \notag\\
&d_{\rm R}^0 =(4-g)(4+3g)/(8g) = 3/4 \,,
\label{eq:CG}
\end{align}
where $g=8/3$ for percolation, in which case the above results are rigorous~\cite{Schramm,KagerCardy}.
The CG duality transformation $g \to 16/g$ relates $d_{\rm H}^0 \to \overline{d}_{\rm H}^0$, 
and leads to the duality relation $(d_{\rm H}^0-1) (\overline{d}_{\rm H}^0-1)=1/4$ \cite{SD87}.

By comparing Eq.~(\ref{eq:CG}) to the numerical results in Tab.~\ref{tab:4},
we obtain that for $n \geq 1$: (1), $d_{\rm F}^{n}$ cannot be described by the exact $d_{\rm F}^{0}$ formula 
in Eq.~(\ref{eq:CG}) that has a minimum $d_{\rm F, min}=(2+\sqrt{3})/2 \approx 1.866$ at $g=2\sqrt{3}$, 
and (2), the recursive clusters violate the domain Markov property, 
since $d_{\rm H}^n$ and $\overline{d}_{\rm H}^n$ data do not satisfy the duality relation.

\paragraph{Limiting clusters.}
The $n$-dependence of critical exponents is illustrated in Fig.~\ref{fig4_dX}(b)-(d).
It is shown that as $n$ increases, the exponents for ${\cal C}_n$ and $\overline{\cal C}_n$
approach each other. They can be convincingly fitted
to ratios of low-degree polynomials, with a common limit for the standard and dense exponents. 
The common limiting values for $n \to \infty$ are estimated as:
\begin{align}
d_{\rm H}^\infty &= \overline{d}_{\rm H}^\infty \simeq 1.42(3) \,, \notag\\
d_{\rm B}^\infty &= \overline{d}_{\rm B}^\infty =
d_{\rm F}^\infty   = \overline{d}_{\rm F}^\infty \simeq 1.826(6) \,, \notag\\
d_{\rm R}^\infty &= \overline{d}_{\rm R}^\infty \simeq 0.00(6) \,.
\label{eq:limdims}
\end{align}
The numerical values of Tab.~\ref{tab:4} also appear to satisfy 
$d_{\rm R}^n = - \overline{d}_{\rm R}^n $ for each $n$; we have no explanation of this.
In particular the common $n \to \infty$ limit $d_{\rm R}^\infty = \overline{d}_{\rm R}^\infty = 0$
most probably holds true.
These results provide substantial evidence that the difference between standard
and dense clusters disappears when $n \to \infty$.

The above results can be used to characterize the limiting clusters
${\cal C}_\infty = \overline{\cal C}_\infty$ in various ways.
The fact that $d_{\rm R}^\infty = 0$ means that the number of red
bonds in the limiting clusters does not grow with $L$. This is compatible with
the observation (Fig.~\ref{fig4_dX}(c)) that the difference between the clusters
and their backbones vanishes in the limit. In other words, the limiting clusters
are dense objects, with only few leaves or dangling ends. Moreover, they are
devoid of deep fjords, since their hulls and external perimeter scale in the
same way (Fig.~\ref{fig4_dX}(b)).

Another set of clusters having similar characteristics are
the FK clusters of the $q=4$ state Potts model,
whose hulls behave as the level lines of a free Gaussian field with central
charge $c=1$. These Potts clusters can be described by the CG construction
with the self-dual choice of the coupling, $g=4$ \cite{loopreview}. 
They have $d_{\rm R} = 0$, $d_{\rm F} = \frac{15}{8}$ and $d_{\rm H} = \frac{3}{2}$, 
coming from Eq.~(\ref{eq:CG}).

Despite of this resemblance, the ${\cal C}_\infty$ clusters are most definitely
different from the $q=4$ FK clusters: 
the fractal dimensions $d_{\rm F}^\infty$ and $d_{\rm H}^\infty$ disagree with Eq.~(\ref{eq:CG}).

\paragraph{Universality.}
Changing the lattice from square to triangular, 
or the process from bond to site percolation~\cite{Lattice}, 
obviously modifies the thresholds $\pc{n}$. 
However, the critical exponents, $d_{\rm F}^n$, $d_{\rm R}^n$ and $d_{\rm H}^n$, 
are found to be unchanged; the critical wrapping probability $R_2^n$ 
also remains the same for different processes on the same lattice. 
This demonstrates the universality of recursive percolation.
Simulations for $d=3$ (see SM) produce different exponents, but the thresholds $\pc{n} < 1$ remain
non-trivial, so we conjecture that recursive percolation is well-defined in any $d$ below the
upper critical dimension $d_{\rm uc} = 6$ \cite{StaufferAharony,Grimmett}.
For $d > d_{\rm uc}$ we expect $\pc{n} = 1$ for $n \ge 1$ and trivial exponents (see SM).

\paragraph{Discussion.}
We introduced a simple lattice model, {\em recursive percolation}, 
which represents an infinite family of new universality classes. 
A crucial element of its definition is that the $n$th recursive process occurs on 
the set of dense percolation clusters $\overline{C}_{n-1}$ such that 
occupied bonds can be placed between those neighboring sites connected via non-local paths.
Indeed, using instead the standard clusters $C_{n-1}$ would have been tantamount to 
a trivial modification of $p$ in the $n=0$ process, leading to $\pc{n} = 1$ for all $n \ge 1$. 
We also stress that although the underlying medium is fractal for $n \ge 1$, 
recursive percolation belongs to the realm of the original $d$-dimensional Euclidean space, 
as witnessed most clearly by the hyperscaling relation in Eq.~(\ref{eq:scaling_P}).
Moreover, we find that the $d=2 $ critical exponents for $n \geq 1$ are beyond 
the description of field-theoretical methods that are applicable for most two-dimensional lattice models.
Several important questions arise:
is recursive percolation conformally invariant at criticality,
what is the universality criterion, 
and how can the exact values of critical exponents be obtained?
Do the $n \geq 1$ clusters enjoy multifractal properties?

It is worth mentioning that earlier studies of statistical models on top of fractal structures \cite{RammalToulouse}
focussed either on the case where the underlying structure is a self-similar set, like the
Sierpi\'nski gasket \cite{Gefen84}, or where the model is random walks of the
self-avoiding (SAW) \cite{RammalToulouseVannimenus,Sahimi84,Vanderzande92,Ferber04,Janke08,Janke14} or loop-erased \cite{Daryaei2014}
types on top of percolation backbones.
While the former case is easy, the latter inherits the difficulties of the underlying $d$-dimensional lattice. 
Interestingly, SAW on backbones defines a new universality class exactly at $p = p_{\rm c}$ \cite{Vanderzande92} 
with multifractal properties \cite{Janke08,Janke14}. 
In a renormalization group  language this means that $p_{\rm c}$ is an unstable fixed point
from which the system may flow to either the usual SAW fixed point at $p=1$, or to a trivial fixed point at $p=0$.
However, recursive percolation differs from these existing works in various ways: 
it can be defined recursively any number of times, and the critical exponents are incompatible with 
existing analytical methods.

We conclude by suggesting that the recursive construction presented here, via the
study of percolation on percolation clusters, may carry over more generally to the
$q$-state Potts model. For instance, it is well known that $q$-state FK
clusters arise by considering percolation with $p_{\rm c} = \sqrt{q}/(1+\sqrt{q})$
on top of $q$-state Potts spin clusters \cite{ChayesMachta}, 
which are widely applied in cluster-type Monte Carlo methods~\cite{SwendsenWang}.
Both types of clusters are well defined for arbitrary real $0 \le q \le 4$ \cite{Dubail10,Vasseur12bis}.
It is thus tempting to speculate that on top of $q$-state FK clusters one may define new
$q_1$-state FK clusters, and that the latter will be critical for a suitable non-trivial
choice of the temperature variable, with distinct critical exponents. Future work
will show whether this construction is possible and can be repeated recursively.

\paragraph{Acknowledgments.}
Two of us (YD and JLJ) thank A.D. Sokal and New York University, where this work was initiated, for hospitality. 
We also thank H.W.J.\ Bl\"ote, D.P. Landau, J.P.\ Lv, J. Machta and R.M. Ziff for valuable discussion.
The research of JLJ was supported by the Agence Nationale de la Recherche
(grant ANR-10-BLAN-0414:~DIME) and the Institut Universitaire de France. The research of YD
was supported by the National Natural Science Foundation of China under Grant No. 11275185
and the Chinese Academy of Sciences. YD also acknowledges the Specialized Research
Fund for the Doctoral Program of Higher Education under Grant No. 20113402110040. 


\begin{thebibliography}{99}
\bibitem{ErrorBar} 
 The effect of uncertainties of $\pc{i}$ ($i=1,\ldots,n$) is not taken into account
 in the error bars quoted in Tab.~\ref{tab:4}. It is  smaller or (at most) comparable to the
 quoted statistical errors.
\end{thebibliography}
\end{document}


\title{Supplemental Material for: Recursive Percolation}

\author{Youjin Deng}
\email{yjdeng@ustc.edu.cn}
\affiliation{Hefei National Laboratory for Physical Sciences at Microscale
and Department of Modern Physics, \\
University of Science and Technology of China, Hefei, Anhui 230026, China}

\author{Jesper Lykke Jacobsen}
\email{jesper.jacobsen@ens.fr}
\affiliation{Laboratoire de Physique Th\'eorique, \'Ecole Normale Sup\'erieure, 24 rue Lhomond,
  75231 Paris, France}
\affiliation{Universit\'e Pierre et Marie Curie, 4 place Jussieu,
  75252 Paris, France}

\author{Xuan-Wen Liu}
\affiliation{Hefei National Laboratory for Physical Sciences at Microscale
and Department of Modern Physics, \\
University of Science and Technology of China, Hefei, Anhui 230026, China}

\maketitle
This supplemental material (SM) provides details on simulations and fitting procedures in both two and three dimensions.

\section{Two Dimensions}

\paragraph{Simulations and Sampled Quantities.}

We investigate recursive percolation on periodic $L\times L$ square lattices. 
To generate a bond configuration for a given $i$th generation, we independently 
visit each edge on the lattice, and randomly place an occupied bond with probability $p^i$ 
if the visited edge connects two sites in the same cluster for the $(i-1)$th generation.
The percolation clusters are then constructed via a breadth-first growing procedure 
similar to the Leath-Alexandrowicz algorithm~\cite{Leath}.
For the $n$th generation of recursive percolation, 
this procedure is performed recursively for $i=0,1,\ldots,n$, 
with the bond probability $p^i$ being set at the previously determined critical value for $i<n$, 
and at any desired value $p$ for $i=n$.

\begin{figure}[htpb]
\centering
\includegraphics[scale=0.35]{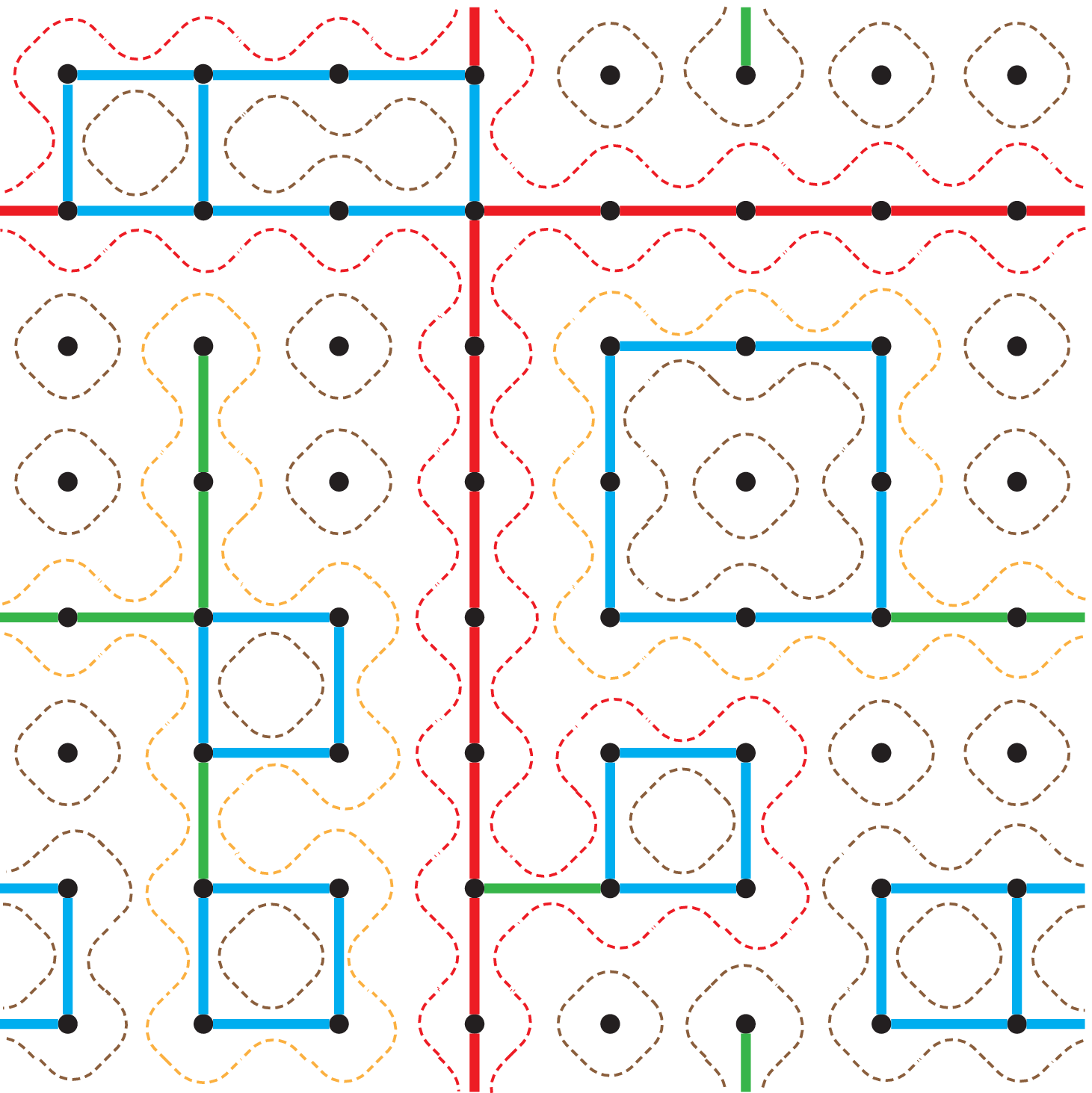}
\caption{A configuration of percolation clusters, of which one contributes to $R_2^n$.
         Bridges are denoted by the green lines, and non-bridges are shown as blue and red bonds
         with the latter for pseudo-bridges.
         The loops separating clusters from their duals are shown as dotted lines on the medial (tilted square) lattice.
}
\label{fig:BKW}
\end{figure}

To characterize the geometric properties of the $n$th generation of percolation clusters, 
the occupied bonds are classified into bridges and non-bridges: 
a bond is a bridge if its deletion leads to the disconnection of a cluster.
A bond configuration is mapped onto the corresponding Baxter-Kelland-Wu (BKW) loop configuration~\cite{BKW76}, 
defined on the medial graph of the square lattice. 
An example is shown in Fig.~\ref{fig:BKW}. 
A pseudo-bridge is defined as a non-bridge but both of whose sides are adjacent to the same loop.
Removing all the bridges leads to a configuration of more compact components, called backbone clusters, 
in which every pair of sites are connected via at least two independent paths.
An efficient algorithm has been introduced in Ref.~\cite{Xiao} for the classification of occupied bonds
into these three classes.

From the simulations of the $n$th generation of recursive percolation, we sampled the following quantities. 
\begin{itemize}
   \item Wrapping probability $R^n_1=\langle r_x+r_y\rangle/2$, with $\langle \cdots \rangle $ for ensemble average. 
   If a configuration connects to itself along the $\alpha$-direction ($\alpha=x,y$) on the lattice,
   we set $r_{\alpha}=1$; otherwise, $r_{\alpha}=0$. 
   \item Wrapping probability along both directions $R^n_2=\langle r_x\cdot r_y\rangle$. 
   \item The  mean density of occupied bonds $\rho^n$.
   \item The mean number of pseudo-bridges $B_{\rm R}^n$.
   \item The mean size of the largest cluster $C^n_1$. 
   \item The mean size of the largest backbone cluster $C^n_{\rm b1}$.
   By definition, the backbone consists of non-bridges.
   \item The mean length of the largest loop $H^n_1$. 
   \item The mean maximum time step $S^n_1$ for constructing percolation clusters. 
         When growing a percolation cluster via the breadth-first method from a seed site, 
         the time step is recorded and measures the shortest-path length between the seed site and any activated site 
         at the current Monte Carlo step~\cite{Zongzheng}. 
         The time step $S$ for a percolation cluster is then the maximum shortest-path length between the seed site 
         and any site in the cluster.
         Given a bond configuration, $S^n_1$ is defined as the maximum time step among all the clusters. 
         
\end{itemize}
The {\em dense} percolation clusters are obtained by filling all the edges that connect two neighboring sites in the same cluster. 
In other words, the dense clusters are the $(n+1)$th generation of percolation clusters with probability $p^{n+1}=1$.
For the dense clusters, the above quantities are measured and denoted by ${\overline R}^n_1$, ${\overline R}^n_2, \ldots$,
with an overline.  By definition, one has $C^n_1 = {\overline C}^n_1$.±

According to finite-size scaling and our previous stu\-dies~\cite{Xiao,Zongzheng}, one expects that at criticality,
\begin{eqnarray}
B_{\rm R}^n &\propto& L^{d^n_{\rm R}} \; , \hspace{5mm} 
C_1^n \propto L^{d^n_{\rm F}} \; , \hspace{5mm} 
C_{\rm b1}^n \propto L^{d^n_{\rm B}} \; , \hspace{5mm} \nonumber \\
H_1^n &\propto& L^{d^n_{\rm H}} \; , \hspace{5mm} 
S_1^n \propto L^{d^n_{\rm S}} \; , 
\label{eq:scaling}
\end{eqnarray}
where $d^n_{\rm R}$ is the dimension of red bonds, $d^n_{\rm F}$ is the cluster's fractal dimension, 
$d^n_{\rm B}$ is the fractal dimension of backbone clusters, $d^n_{\rm H}$ is the hull dimension, and 
$d^n_{\rm S}$ is the shortest-path dimension. 

For each generation $n$, we ran preliminary simulations within a relatively wide range of $p^n$
and for relatively small values of $L$, and obtained a rough estimate $\pc{n}$
by studying the finite-size scaling of dimensionless obser\-vables.
Simulations were then performed at and near the value of $\pc{n}$ estimated in the initial runs.
This procedure was iterated a number of times,
and finally we simulated at the best estimate $\pc{n}$ to determine several sets of critical exponents.
The number of samples (in units of $10^9$) is
about 6.0, 6.0, 5.0, 3.0, and 3.0 for $n=1,2,\ldots,5$, respectively.

\begin{figure}[htpb]
\centering
\includegraphics[scale=0.65]{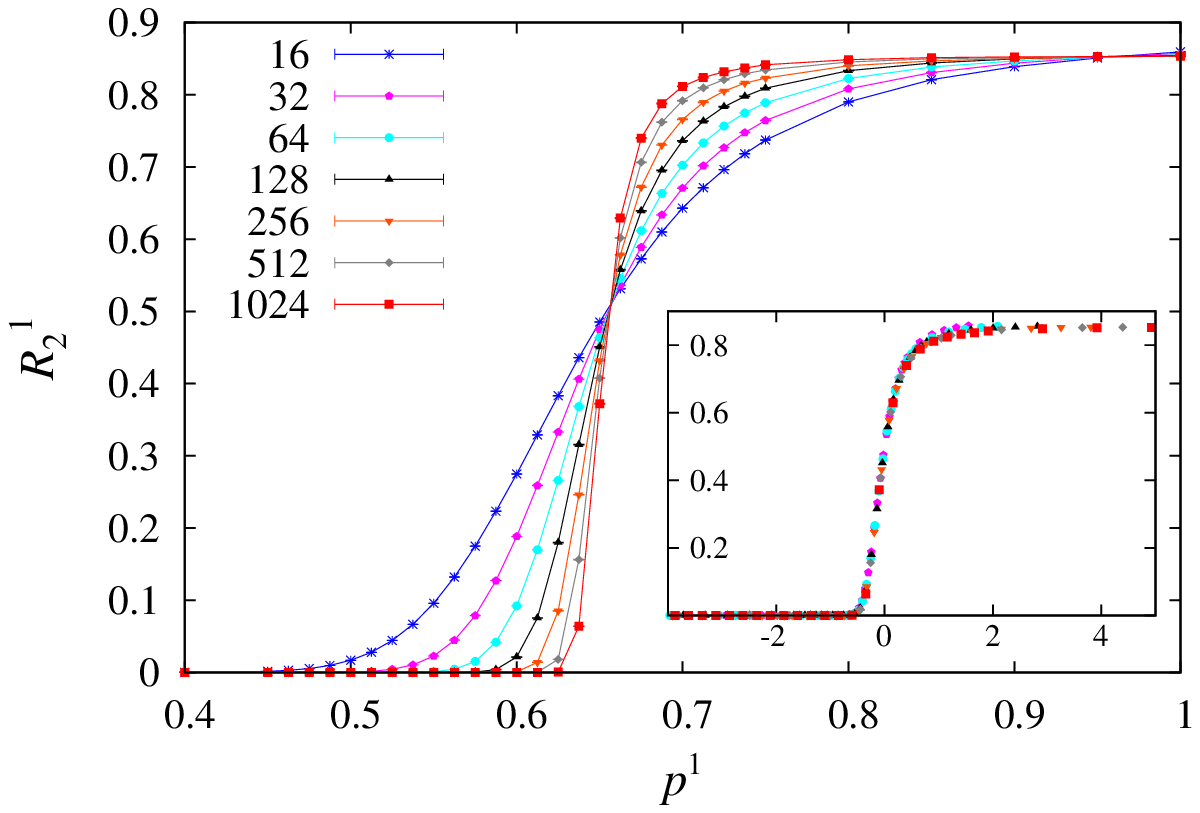}
\caption{Wrapping probability $R_2^1$ versus bond probability $p^1$ for square lattices of different sizes $L$.
         The inset shows the $R_2^1$ data as a function of $u = (p^1-\pc{1}) L^{y_t^1}$, 
         where $\pc{1}=0.654902$ and $y_t^1=0.433$ are taken from the fit.}
\label{fig:p_n1}
\end{figure}

\paragraph{Wrapping probability.}

The wrapping probabilities, universal at criticality, are known to be powerful in loca\-ting a phase transition. 
Indeed, the non-trivial percolation threshold in the $n=1$ generation is clearly
demonstrated by Fig.~\ref{fig:p_n1}, which plots the $R_2^1$ data versus bond probability $p^1$. 
The approximately common intersection for different sizes yields an approximate threshold $\pc{1} \approx 0.65$.
For $p^1 < \pc{1}$, the wrapping probability $R_2^1$ quickly drops to zero, 
illustrating a non-percolating sub-critical phase.
For $p^1 > \pc{1}$, as $L$ increases, $R_2^1$ converges to a non-trivial value $R_{\rm 2, d}^1 \approx 0.85$. 
This implies that for $p^1 > \pc{1}$ , the system is in a quasi-long-range ordered phase within the same 
universality class---namely, the two-point correlation function decays algebraically 
as a function of distance with a $p^1$-independent exponent.
In the terminology of the renormalization group, one has an unstable fixed point at $p^1 = \pc{1}$, 
a trivial fixed point $p^1 = 0$, 
and a non-trivial fixed point  at $p^1 \approx 1$ 
(the second intersection at about $p^1 \approx 0.95$ for small sizes may move to the right for larger systems).
It is interesting to observe that using $u = (p^1-p_c^1) L^{y_t^1}$ with $y_t^1 = d_{\rm R}^1 = 0.433$ 
and $p_c^1 = 0.654902$ from the fitting result, 
the $R_2$ data for different sizes approximately collapse onto a single curve, as shown by the inset of Fig.~\ref{fig:p_n1}.

\begin{figure}[htpb]
\centering
\includegraphics[scale=0.6]{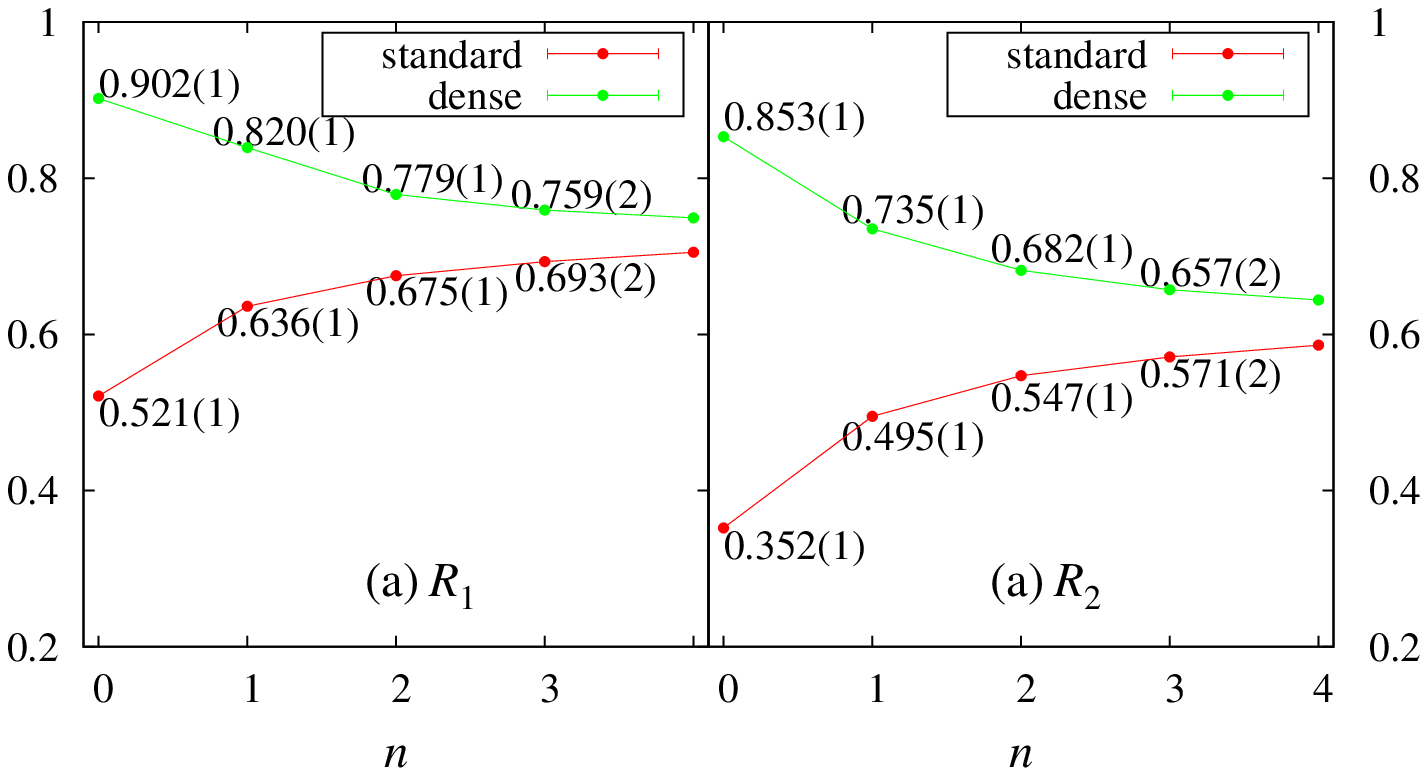}
\caption{The wrapping probabilities $R^n_1$ (${\overline R}^n_1$) and $R^n_2$ (${\overline R}^n_2$) are shown in frames (a) and (b) respectively as functions of $n$ at the percolation threshold $p_c^n$.}
\label{fig:wrap}
\end{figure}

The plots of $R_2^n$ against $p^n$ near $\pc{n}$ (see Fig. 1 in the manuscript) 
present very clear crossings for $n=1,2,3,4$, but starting from $n=5$ the scaling becomes more problematic
in the sense that larger sizes $L$ are needed to see a clearer crossing.
The critical thresholds $\pc{n}$ are then determined by least-squares fitting of the $R_2$ (and $R_1$)
data near $\pc{n}$ by 
\begin{align}
\mathcal{O}^n &= \mathcal{O}^n_{\rm c} +a_1\left(p^n-\pc{n}\right)L^{y_t^n} +a_2\left(p^n-\pc{n}\right)^2L^{2y_t^n} \notag \\
      & +b_1L^{y_1^n} +c \;  \left(p^n-\pc{n}\right) L^{y_t^n+y_1^n}+ b_2 L^{-2} + \cdots \,,
\label{eq:2.1}
\end{align}
defining the thermal exponent $y_t$ and the first correction to scaling exponent $y_1$.
The coefficients $a_1$, $a_2$, $b_1$, $b_2$, and $c$ are non-universal constants.
The correction term with exponent $-2$ is generally expected to be present in two dimensions.
We have determined the thresholds $\pc{n}$ for $n \le 5$, and 
the leading correction exponent $y_1^n$ is found around $-1$, independently of $n$.
The error bars in Tab. 1 in the manuscript include the uncertainties
in estimating $y_1$, excluding/including some terms in Eq.~(\ref{eq:2.1}),
and varying the minimum size for a lower cutoff $L \geq L_{\rm min}$.
The critical values $R_{1,c}^n$ and $R_{2,c}^n$ are shown in Fig.~\ref{fig:wrap},
where the ``dense" wrapping probabilities $\overline{R}_1$ and $\overline{R}_2$ are obtained from the ansatz 
\begin{equation}
\mathcal{O}(L) = \mathcal{O}_c + b_1 L^{-1} +b_2 L^{-2} \; ,
\end{equation}
with fitting parameters $b_1$ and $b_2$.

Several interesting features are implied by Fig.~\ref{fig:wrap}. 
For the standard bond percolation $n=0$,  when we densify critical bond percolation clusters  by adding bonds
between all pairs of neighboring sites in the same cluster, 
the wrapping probabilities increase significantly---$R^0_1 = 0.521 \rightarrow {\overline R}^0_1 = 0.902$ 
and $R^0_2 = 0.352 \rightarrow {\overline R}^0_2 = 0.853$.
For the wrapping probability along either of the two lattice directions $ R^0_e = 2 R^0_1- R^0_2$ , one has 
$R^0_e = 0.690 \rightarrow {\overline R}^0_e = 0.951$; namely, only less than 5\% of the critical dense percolation clusters do not percolate.
This can be understood from the fact that as the percolation point is approached, 
large clusters are joined together by a few bonds, and that they often need just one bond to become wrapping.
 
As the generation number $n$ becomes larger,
the critical wrapping probabilities of the ``standard" recursive percolation clusters increase 
while the dense ones decrease.
The difference between the ``standard" and the ``dense" wrapping probabilities becomes smaller and smaller 
and probably vanishes for $ n \rightarrow \infty$.

\paragraph{Scaling function.}
The probability distribution $P(s,L)$ for the size of the $n=1$ critical cluster size, 
obtained in the single-cluster growing procedure, is shown in Fig. 3 in the manuscript. 
This plot, together with the two insets, strongly support that $P(s,L)$ for $n=1$ enjoys the standard scaling form 
given by Eq.~(2) and the hyperscaling relation $\tau=1+d/d_{\rm F}^1$ involves the original spatial dimensionality $d=2$. 
The blue and red curves in the main plot exhibit an almost perfect power law behavior over six decades. 
The difference in their slopes is slight, but can be detected from the figure (the blue curve is above the red one for small $s$, and vice versa).
To magnify the difference between the 0th and 1st generations, the upper inset partially ``suppresses'' the algebraic behavior 
by setting the vertical axis to be $s P(s,L)$. 
To display the universal function $f(s/L^{d_{\rm F}})$ in Eq. (2), the lower inset completely ``eliminates" the algebraic behavior 
by plotting $s^{\tau-1} P(s,L)$ versus $s/L^{d_{\rm F}}$ for different sizes $L$. 

Further, the three distinct behaviors of $P(s,L)$ are  displayed by the blue curve in the upper inset. 
For small cluster sizes $s \sim $O(1) of order of lattice spacing,  the probability distribution is non-universal 
(the correction exponent may be universal).
For size $1 \ll s \ll L^{d_{\rm F}}$, the function $f(s/L^{d_{\rm F}}) \approx f(0) $ and thus  $P(s,L) \sim s^{1 -\tau}$, independently of $L$.
For  $s \sim L^{d_{\rm F}}$,  $f(s/L^{d_{\rm F}})$ becomes important and are responsible for the bump and the dip of $P(s,L)$ in Fig. 3.
For $s > L^{d_{\rm F}}$, the cluster size is limited by the system volume and $P(s,L)$ drops super-exponentially.

\begin{table}[htpb]
\centering
\caption{Bond density $\rho^n$ and $\overline{\rho}^n$ for standard and dense clusters.
 For $n=0$, $\overline{\rho}^n=3/4$ can be exactly derived~\cite{Hu2014}.}
\begin{tabular}{llllll}
\hline
$n$ & 0 & 1 & 2 & 3 & 4 \\
\hline
$\rho^n$ & 0.5  & 0.491(1) & 0.489(1) & 0.491(1) & 0.494(1) \\
$\overline{\rho}^n$ & 0.75 & 0.661(1) & 0.618(1) & 0.592(1) & 0.577(1) \\
$\frac{n+3}{2(n+2)}$ & 0.75 & 0.667 & 0.625 & 0.6 & 0.583 \\
\hline
\end{tabular}
\label{tab:2}
\end{table}

\paragraph{Bond density.}
The density of occupied bonds for the standard and the dense clusters, 
$\rho^n$ and $\overline{\rho}^n$, is given in Tab.~\ref{tab:2}.
The results are close, albeit not perfectly equal, to $\rho^n = \frac12$ independently of $n$, respectively
$\overline{\rho}^n = \frac{n+3}{2(n+2)}$, with the common limit
$\rho^n,\overline{\rho}^n \to \frac12$ as $n \to \infty$. 

We conjecture that $\rho^\infty = \overline{\rho}^\infty = \frac12$ holds exactly. 
Given a configuration of occupied bonds, one can perform a duality mapping and 
construct the associated percolation clusters on the dual lattice. 
Note that the standard and the ``dual" clusters share the same BKW loop configuration
and have the same hull exponent.
In the $n \rightarrow \infty$ limit, the occupation probability 
$\pc{\infty} \rightarrow 1$ implies that the two ending sites of an empty edge must be
in different clusters, and that the resulting ``dual" clusters reduce to a set of backbone clusters.
As mentioned in the manuscript, the standard clusters in the limit are also compact: 
the difference from the backbone clusters vanishes, 
the hulls and external perimeters scale in the same way,
and the number of red bonds does not increase with $L$.
Therefore, we argue that for $n \rightarrow \infty$, 
the standard and the ``dual'' clusters become identical, 
and that the bond density is $\rho^\infty = 1/2$.

\paragraph{Critical exponents.}
To determine the critical exponents in Eq.~(\ref{eq:scaling}), we carry out simulations 
at the estimated thresholds $p_c^n$ for $n=1,2,3,4$.
The Monte Carlo data for the quantities in Eq.~(\ref{eq:scaling}) are analyzed by
\begin{equation}
\mathcal{O}(L) = L^{d_\mathcal{O}} (b_0 + b_1 L^{-1} +b_2 L^{-2}) \; ,
\label{eq:1}
\end{equation}
where $\mathcal{O}$ may represent, respectively, the number of pseudo-bridges $B_{\rm R}^n$, 
the largest cluster size $C_1^n$, the size of the largest backbone clusters $C_{\rm b1}^n$, 
the largest loop length $H_1^n$, and the maximum construction time step $S_1^n$.

\begin{table}[htpb]
\centering
\caption{Values of shortest-path exponents. 
 For $n=0$, it has been measured as $d_{\rm S}^0 = 1.130 \, 77(2)$ \cite{Zongzheng}.}
\begin{tabular}{llllll}
\hline
$n$                         &\;0         &\;1         &\;2          &\;3          &\;4 \\
\hline
$d_{\rm S}^n$               &\;1.1308(2) &\;1.0813(2) &\;1.0645(10) &\;1.0576(10) &\;1.053(1) \\
$\overline{d}_{\rm S}^n$    &\;1.0061(2) &\;1.0212(5) &\;1.031(1)   &\;1.036(1)   &\;1.039(1) \\
\hline
\end{tabular}
\label{tab:short_path}
\end{table}

The estimates for the shortest-path exponents, $d_{\rm S}^n$ and $\overline{d}_{\rm S}^n$, 
are shown in Tab.~\ref{tab:short_path}, 
and the results for the other exponents are summarized in Tab. II and shown in Fig.~4 in the manuscript. 
As $n$ increases, it can be seen that the difference in each pair of exponents decreases, 
namely ($d_{\rm H}^n$, $\overline{d}_{\rm H}^n$), ($d_{\rm F}^n$, $d_{\rm B}^n$), 
($d_{\rm S}^n$, $\overline{d}_{\rm S}^n$), and ($d_{\rm R}^n$, $\overline{d}_{\rm R}^n$).
This suggests, once again, that as $n$ increases, the distinction between the standard and the dense percolation clusters disappears.
Further, the decrease of $d_{\rm R}^n$ implies that the percolation clusters 
become more and more compact. 

\begin{figure}[htpb]
\centering
\vspace{0.5cm}
\hspace*{-1.1cm}\includegraphics[scale=0.75]{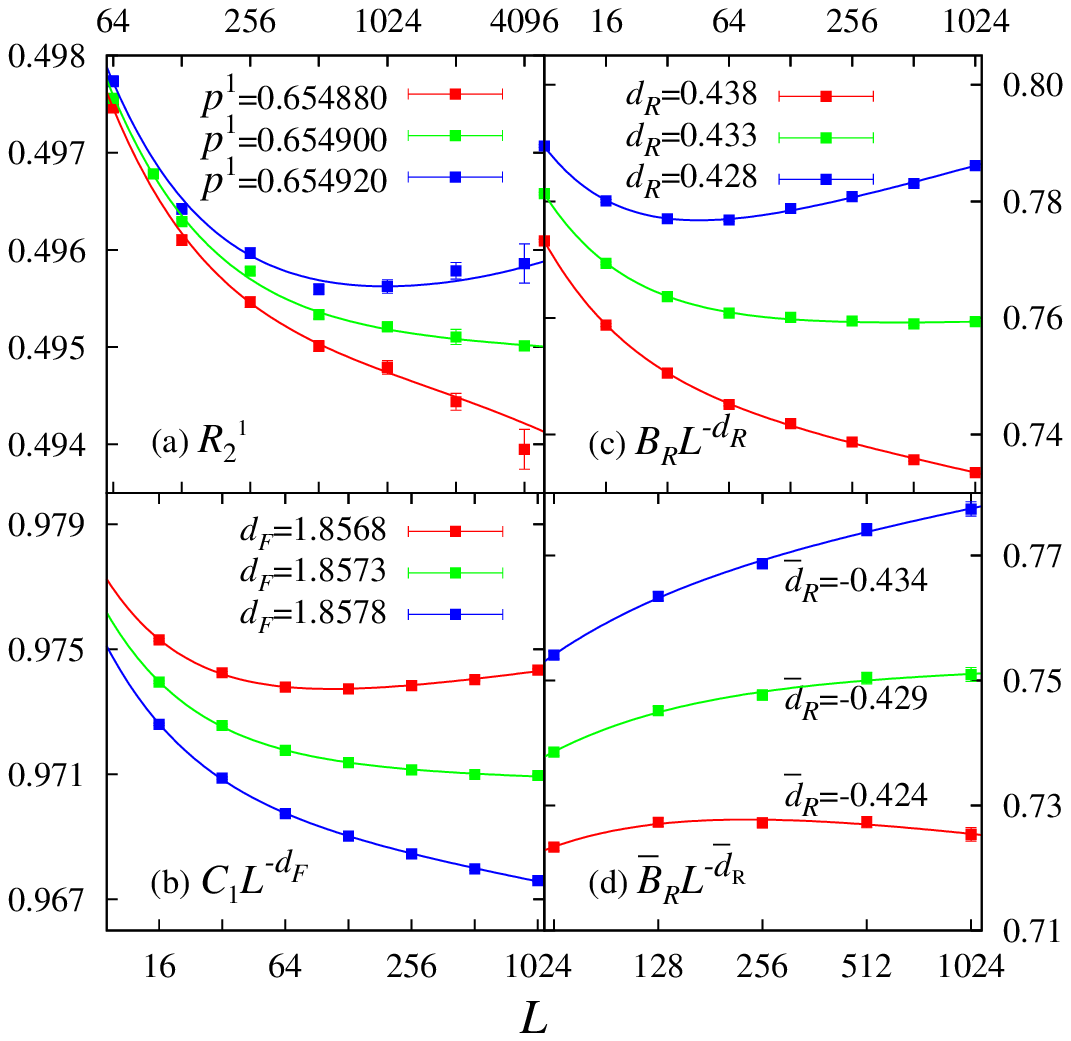}
\vspace{1cm}
\caption{Test of the estimated error bars. 
      (a), the wrapping probability $R_2^1$ versus linear size $L$ at $p^1=0.65488, 0.65490$ and $0.65492$, with an interval of two reported error bars for $p_c^1$.
     (b-d), scaled observable $\mathcal{O} L^{-d_\mathcal{O}}$ versus $L$ for $\mathcal{O}(L)=C_1, B_{\rm R}$ and $\overline{B}_{\rm R}$, respectively. 
     The interval of the $d_\mathcal{O}$ values for different curves correspond to five reported error bars. 
      The curves are obtained from the fits.}
\label{fig:err}
\end{figure}

\begin{table}[htpb]
 \centering
 \caption{Fits of critical exponents at $n=\infty$.}
\scalebox{1.0}{
 \begin{tabular}{Ic|c|l|l|l|cI}
 \whline
 $d_{\mathcal{O}}$                    & $n_{min}$ & ~~~$a_{1}$     & ~~~$a_{2}$    & ~~~$d_{\mathcal{O}}^{\infty}$ & $\chi^2$/DF \\
 \whline
 $d_H$                                &  2  & ~0.43(1)    & -0.23(1)  & ~1.380(2)      & 0/0\\
 \hline
 {\multirow{2}{*}{${\overline d}_H$}} &  2  & -0.0583(8)  & ~~~~-     & ~1.4221(4)     & 0/1\\
  \cline{2-6}
	                                    &  1  & -0.080(2)   & ~0.028(1) & ~1.4258(5)     & 2/1\\
 \whline
 {\multirow{2}{*}{$d_F$}}             &  2  & ~0.0403(8)  & ~~~~-     & ~1.8222(4)     & 0/1\\
  \cline{2-6}
                                      &  1  & ~0.051(2)   & -0.014(2) & ~1.8203(5)     & 0/1\\
 \hline
 {\multirow{2}{*}{$d_B$}}             &  2  & -0.0822(8)  & ~~~~-     & ~1.8353(4)     & 0/1\\
  \cline{2-6}
                                      &  1  & -0.095(2)   & ~0.018(2) & ~1.8377(5)     & 0/1\\
 \whline
 {\multirow{2}{*}{$d_R$}}             &  1  & ~0.87(3)    & -0.37(2)  & -0.070(8)      & 3/1\\
  \cline{2-6}
	                                    &  2  & ~1.2(2)     & -0.8(2)   & -0.13(3)       & 0/0\\
 \hline
 {\multirow{2}{*}{${\overline d}_R$}} &  2  & -0.49(3)    & ~~~~-     & -0.03(1)       & 0/1\\
  \cline{2-6}
	                                    &  1  & -0.70(6)    & ~0.26(4)  & -0.01(2)       & 0/1\\
 \whline
 {\multirow{2}{*}{$d_S$}}             &  2  & ~0.045(6)   & ~~~~-     & ~1.042(2)      & 0/1\\
  \cline{2-6}
	                                    &  1  & ~0.06(1)    & -0.017(9) & ~1.040(3)      & 0/1\\
 \hline
 {\multirow{2}{*}{${\overline d}_S$}} &  2  & -0.032(5)   & ~~~~-     & ~1.047(2)      & 0/1\\
  \cline{2-6}
	                                    &  1  & -0..04(1)   & -0.017(9) & ~1.049(3)      & 0/1\\
 \whline
 \end{tabular}
}
 \label{Tab:fit}
 \end{table}

To see whether the ``standard" and ``dense" critical exponents have a common limiting value for $n \rightarrow \infty$, 
we simply apply a low-degree polynomial ansatz
\begin{equation}
d_\mathcal{O}^n=d_\mathcal{O}^{\infty}+a_{1}\frac{1}{n}+a_{2}\frac{1}{n^2} \; .
\label{eq:3}
\end{equation}
The fitting results are summarized in Tab.~\ref{Tab:fit}. 
There is an unphysical artifact: the backbone dimension $d_{\rm B}^\infty$ of percolation clusters 
is bigger than the corresponding fractal dimension $d_{\rm F}^\infty$, 
and the hull dimension $d_{\rm H}^\infty$ is smaller than the dense one $\overline{d}_{\rm H}^\infty$.
Actually, the estimate $d_{\rm H}^\infty$ is already smaller than $\overline{d}_{\rm H}^n$ for $n=2,3,4$ 
(see Tab.~II in the manuscript).
This is due to the fact that we have only a limited number of data points and the ansatz~(\ref{eq:3}) is pro\-bably oversimplified.
On the other hand, this supports that for $n \rightarrow \infty$, the distinction
between the standard and the dense clusters vanishes, and the convergence is faster than $1/n$.
On the basis of these fitting results, we estimate the limiting values of the critical exponents as in Eq.~(3) of the main manuscript.

To check the estimated error bars in the manuscript and the supplemental material, 
we carried out additional simulations for the $n=1$ recursive percolation at $p^1= 0.654880$ and $0.654920$, 
about two error bars away from $p_c^1 =0.654902$. 
The data are plotted in Fig.~\ref{fig:err}(a). 
As $L$ increases, the middle curve asymptotically becomes horizontal, supporting $p_c^1 \sim p^1=0.654900$.  
In contrast, the upper (lower) curves start to bend upward (downward) for large $L$, suggesting that they are already 
away from the percolation threshold $p_c^1$. 
Taking into account that such an eye-view fitting is very crude, we conclude that the estimate $p_c^1 = 0.654902(10)$ has a very conservative error bar. 
Similarly, we plot in Fig.~\ref{fig:err}(b-d) the scaled observable $\mathcal{O} L^{-d_\mathcal{O}}$ versus $L$ for 
$\mathcal{O} = C_1, B_{\rm R}$, and $\overline{B}_{\rm R}$, respectively. 
The values of $d_\mathcal{O}$ for the middle curve, which are respectively $d_{\rm F}$, $d_{\rm R}$ and $\overline{d}_{\rm R}$, 
are from Table II in the manuscript;  and those for the upper and lower curves are five error bars away the middle one. 
These plots demonstrate the reliability of the results and the associated error bars reported in this work.

\section{Three and higher Dimensions}

\begin{figure}[htpb]
\centering
\includegraphics[scale=0.7]{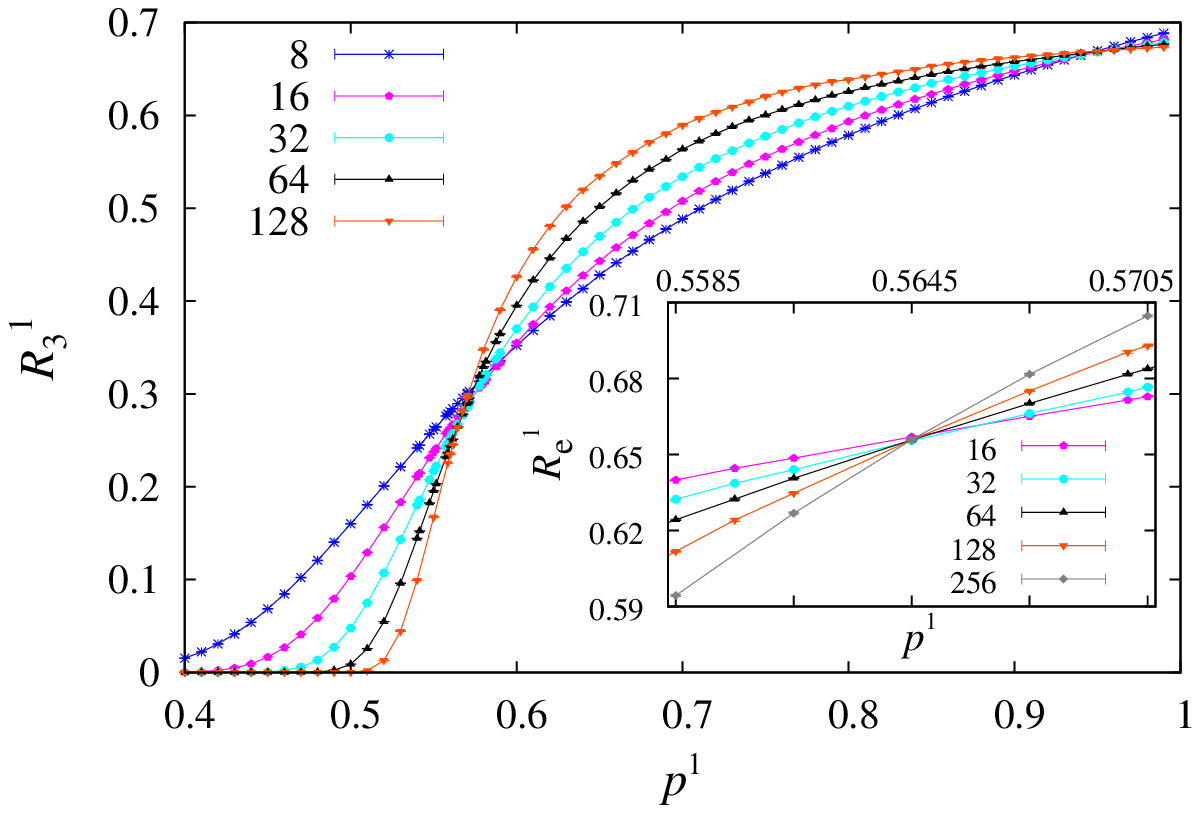}
\caption{Wrapping probability $R_3^1$ as a function of $p^1$ for different sizes $L$ on the simple cubic lattice. 
         The inset is for the wrapping probability $R_e^1$ along either of the three directions.}
\label{fig:R3}
\end{figure}

We have also simulated the recursive percolation model on the $L \times L \times L$ simple-cubic lattice for generation $n=1$. 
The system size is taken as $L=8, 16, \ldots, 256$, and periodic boundary condition are applied.
The total number of samples is about $10^9$.
Figure~\ref{fig:R3} shows the wrapping probability $R_3^1$ (the probability that there exists at least 
one cluster wrapping all three periodic lattice directions) 
as a function of $p^1$ for different sizes. 
As for the two-dimensional case, it is clearly demonstrated that there exists a percolation threshold $p_c^1$ 
separating the non-percolating phase for $p^1<p_c^1$ and the critical phase for $p^1>p_c^1$.
The crossing for small sizes suggests the existence of an additional non-trivial fixed point at $p^1\approx 0.95$.
Nevertheless, we cannot exclude that this is due to some combined effect of various correction terms, 
and the actual fixed point may go to  $p^1 = 1$ if larger sizes were considered.
The clarification of this point would request accurate numerical data, 
since the curves for $p^1>p_c^1$ become close to horizontal for large $L$; this is a generic property of stable fixed points. 
The inset of Fig.~\ref{fig:R3} is for the wrapping probability $R_e^1$ along either of the three directions, 
and yields a rough estimate of the percolation threshold $p_c^1 \approx 0.5645$.
The least-squares fit of the $R_e^1$ data by Eq.~(1) in the manuscript gives 
$p_c^1=0.5647(2)$ and a thermal exponent $y_t^1=0.432(8)$.
It is interesting to observe that, within the error bars, the critical exponents $d_{\rm R}^1 \equiv y_t^1$ 
in two and three dimensions are identical.

Above the upper critical dimension, $d > d_{\rm uc} = 6$, standard percolation has mean-field exponents 
described by the Gaussian fixed point of $\phi^3$ field theory. The percolation threshold $\pc{0} \propto 1/d$,
and more precisely there is a systematic expansion of $\pc{0}$ in powers of $s = 1/(2d - 1)$ \cite{Grassberger03}. At criticality, the
construction of a percolation cluster is basically a branching process with small corrections. This results in tree-like structures,
and since $d_{\rm H} < d$ in mean-field theory, fjords should be very rare. We therefore expect that the densification process,
${\cal C}_0 \to \overline{\cal C}_0$, that forms part of the definition of $n=1$ recursive percolation will only very rarely lead
to the formation of loops. It follows that $\pc{1} = 1$. Our numerical data for $\pc{0}$ and $\pc{1}$
in $d=2$ and $d=3$ exhibit a trend in line with this scenario---($p_c^0=1/2$, $p_c^1=0.654 \, 902$) for $d=2$
while ($p_c^0=0.248 \, 811 \, 8$, $p_c^1=0.5645$) for $d=3$.

By this argument, recursive percolation for any $n \ge 1$ should coincide with standard $n=0$ percolation, 
and the novel physics of the recursive percolation arises from strong critical fluctuations in low dimensions.
We leave the explicit numerical check of this statement for $d > d_{\rm uc}$ to future work.